\documentclass[aps, prd, reprint, nofootinbib, a4paper, 10pt]{revtex4-2}
\usepackage{graphicx}
\usepackage{dcolumn}
\usepackage{hyperref}
\usepackage[english]{babel}
\usepackage[autostyle, english = american]{csquotes}
\MakeOuterQuote{"}

\usepackage{physics}
\usepackage{xcolor}

\DeclareMathOperator\supp{supp}

\usepackage[T1]{fontenc}
\usepackage[scaled, mono = false]{libertine}
\usepackage[libertine, frenchmath]{newtxmath}

\usepackage{amsmath}
\usepackage[scr = euler, cal = cm]{mathalpha}

\begin{document}

\title{Entropy for spherically symmetric, dynamical black holes from the relative entropy between coherent states of a scalar quantum field}
\author{Edoardo D'Angelo }
\email{edoardo.dangelo@edu.unige.it}
\affiliation{Dipartimento di Matematica, Università di Genova - Via Dodecaneso 35, I-16146 Genova, Italy \\ Istituto Nazionale di Fisica Nucleare, Sezione di Genova - Via Dodecaneso 33, I-16146, Genova, Italy}

\begin{abstract}
The goal of this paper is to prove an area law for the entropy of dynamical, spherically symmetric black holes from the relative entropy between coherent states of the quantum matter, generalising the results by Hollands and Ishibashi \cite{HollandsIshibashi19}. We consider the relative entropy between a coherent state and a suitably chosen asymptotically vacuum state for a scalar quantum field theory propagating over a dynamical black hole. We use the conservation law associated to the Kodama vector field in spherically symmetric spacetimes, and the recent results on the relative entropy between coherent states found in \cite{Longo19, CasiniGrilloPontiello19}. We consider the back-reaction of the quantum matter on the metric. From the conservation law associated with the Kodama vector field, we obtain an equation in the form $(S + A/4)' =\Phi$, where $S$ is the relative entropy between coherent states of the scalar field, $A$ is the apparent horizon area, and $\Phi$ is the flux radiated at infinity. The prime denotes a derivative along the outgoing light-rays.
\end{abstract}
\maketitle

\section{Introduction. A quantum cup of tea} 
\label{sec:intro}
The idea to assign an entropy dates back to the beginning of the '70s, when John Wheeler and his graduate student, Jacob Bekenstein, were discussing over a cup of tea \cite{Oppenheim15} the idea that black holes can be described by a handful of parameters only: in their own words, that black holes have "no hair" \cite{Misner74}. "What would happen", asked the professor, "if I throw this cup of tea into a black hole?"

General Relativity's answer is that black holes do not conserve any trace of the cup of tea's microscopic properties. This is however in contradiction with the second law Thermodynamics, because, following this idea, the entropy of the Universe would decrease every time an object falls behind the black hole horizon.

Indeed, the concept of black hole entropy had already been introduced in the context of a formal mathematical analogy between the laws of black hole mechanics and the laws of thermodynamics \cite{Bardeen73}. However, Bekenstein proposed \cite{Bekenstein73} that the black hole horizon area should be understood as a true measure of the information lost into the black hole, and based on information-theoretic arguments he predicted a proportionality factor between entropy and area of $\eta \simeq \frac{1}{2}\ln 2 $.

Bekenstein's theoretical insight was motivated by a consistency requirement between Thermodynamics and General Relativity (GR) only; in 1974, Hawking discovered the physical process by which black holes actually radiate \cite{Hawking74BHExplosions}. His idea has been to substitute the cup of tea with a \textit{quantum cup of tea}; that is, to consider the effects of a black hole on the propagation of a quantum field. He showed that black holes do indeed emit a flux of particles towards a distant observer with a black-body spectrum, at a temperature $T = 1/4m$, where $m$ is the black hole's mass \cite{Hawking74}. Using the analogy between black hole mechanics and thermodynamics, he could then fix the proportionality factor between entropy and area to $\eta = 1/4$, obtaining the celebrated formula
\begin{equation}
S_{BH} = \frac{A_{\mathscr H}}{4} \ , 
\end{equation}
where $A_{\mathscr H}$ is the horizon area of the black hole, and $BH$ can stand for \textit{Black Hole} or \textit{Bekenstein-Hawking}, according to taste.

The question then became how the Bekenstein-Hawking entropy could emerge from a statistical measure of the microscopic degrees of freedom of the black hole. The first proposal in this sense has been made by Bombelli and his collaborators \cite{Bombelli86}: their idea was that the black hole entropy came from the entanglement entropy between the quantum degrees of freedom inside and outside the black hole event horizon. The entanglement entropy was computed as a von Neumann entropy for the reduced density matrix, obtained tracing over the degrees of freedom outside the event horizon. They found that the entanglement entropy is indeed proportional to the spatial area of the entangling surface, but it is divergent in the continuum limit, with a leading term that is regularisation-dependent, and also depends on the number of fields present in the model (see e.g. \cite{HollandsSanders17} for a review on the entanglement measures in the algebraic approach to QFT, and \cite{Solodukhin11} for the applications to black hole entropy).

In this paper, we shall prove an area law for the entropy of dynamical black holes, computing the variation of the entropy of a quantum field propagating over a background metric, and the associated back-reaction on the black hole horizon area. The approach follows closely the original ideas of Bekenstein, Hawking and Bombelli and his collaborators, and generalises the result of \cite{HollandsIshibashi19} for the entropy of a Schwarzschild black hole to the case of dynamical spacetimes.
 
We therefore consider a spherically symmetric, asymptotically flat black hole. On the black hole background, we consider the propagation of a free, massless, scalar field. The key difference from \cite{Bombelli86} relies on the choice of the \textit{relative entropy} as the entropy measure, defined via the Araki formula as the expectation value of a particular operator, constructed using the Tomita-Takesaki relative modular operator \cite{Haag67, Takesaki70}. The relative entropy measures the distinguishability between two states \cite{CasiniGrilloPontiello19}, and, contrary to the case of entanglement entropy, it is directly defined for continuum theories such as QFT. In the special case of Quantum Mechanics, it reduces to the entanglement entropy, regularised subtracting a vacuum term.

We address the problem from the point of view of Quantum Field Theory on Curved Spacetimes (QFTCS) formulated in the algebraic approach \cite{AAQFT15}. In this formalism, a quantum field is considered as the abstract generator of a unital $*$-algebra $\mathcal A$, satisfying a set of relations encoding the canonical commutation rules, the equations of motion, and linearity. On curved spacetimes, the choice of a vacuum state can be non-unique (see e.g. \cite{Candelas80}), and the algebraic approach lets one discuss the properties of the observables without referring to a particular state. The choice of a physical state as a positive, normalised functional $\omega : \mathcal A \to \mathbb C$ then lets one compute the expectation values of observables, and lets one recover the usual representation of the algebra as linear operators acting on a Hilbert space via the Gelfand-Naimark-Segal (GNS) theorem \cite{Haag92}.

Using the algebraic formalism, in \cite{Longo19, CasiniGrilloPontiello19} it has been shown that the relative entropy between the vacuum and a coherent perturbation is given by the integral on a 3-surface of the stress-energy tensor of the coherent perturbation.

In order to generalise their result to a black hole setting, we choose a ground state for the quantum field which is the vacuum with respect to modes coming from past infinity, and regular (in Hadamard sense \cite{Kay88}) otherwise. Our model is then independent on the details of the collapse of the black hole. We then consider a classical, massless wave, which acts as a perturbation on the geometry and as a coherent state with respect to the ground state, switched on at the advanced time $v = v_0$.
 
\begin{figure}
\centering
\includegraphics[scale=0.35]{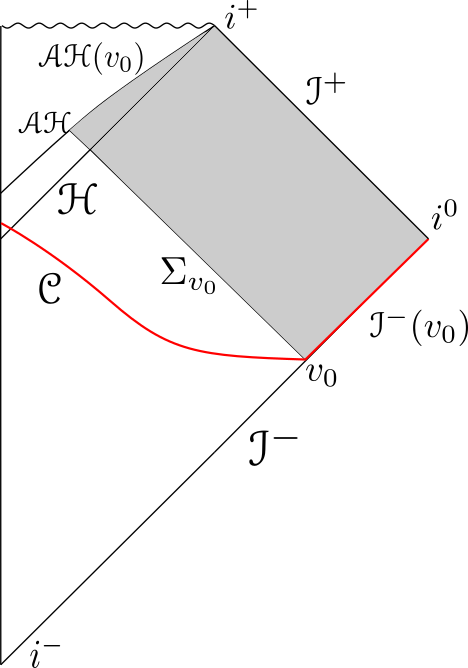} 
\caption{\small{Penrose diagram of the geometric background we are interested in, a dynamical black hole evolving due to a non-vanishing energy-momentum tensor. The black hole develops an apparent horizon $\mathscr{AH}$ inside the event horizon $\mathscr H$, which evolves according to the presence of matter. The initial data surface $\mathscr C$ for the propagation of the quantum field is in red; we give non-vanishing initial data for the perturbation on the segment lying at past null infinity. To prove our result, we shall consider the flux of the Kodama current across the four boundaries of the shaded region.}}
\label{diagram}
\end{figure}
 
Since the space is assumed to approach a Schwarzschild background in the asymptotic past, we can use the same formula as in \cite{HollandsIshibashi19} for the relative entropy between the coherent state and the ground state, which is given in terms of an integral on $\mathscr I^-(v_0)$ (the segment of past null infinity with $v \geq v_0$) of the stress-energy tensor of the coherent wave. The formula is given in \eqref{eq:EE-foundamental-formula}.

The key idea of the paper is that the relative entropy formula can be written as the flux integral of a conserved current at past null infinity. To this end, we make use of the Kodama vector field \cite{Kodama79}. In spherically symmetric spacetimes, the Kodama vector takes a role analogous to a time-like Killing vector in static spacetimes, and in particular its contraction with a conserved stress-energy tensor gives rise to a conserved current. We show that the derivative of the relative entropy along the geodesics tangent to past null infinity equals the flux of the Kodama current on $\mathscr I^-$.

The last step is to connect the variation of the relative entropy to a variation of the horizon area, which evolves both because of the dynamics of the background matter, and of the back-reaction of the coherent wave on the metric via the Einstein field equations.  For the boundary of the dynamical black hole, we consider the local definition of apparent horizon provided by Hayward \cite{Hayward93}. We then consider the flux of the Kodama current across the closed boundary of a region $\mathscr O$ outside the black hole, shown in grey in figure \ref{diagram}. The boundary is naturally composed by the null hypersurface $\Sigma_{v_0}$ at $v = v_0$, the future null, time-like, and space-like infinity $\mathscr I^+$, $i^+$, and $i^0$, $\mathscr I^-(v_0)$, and the corresponding part of the perturbed apparent horizon with $v \geq v_0$, denoted $\mathscr{AH}(v_0)$. Since the Kodama current is conserved, the flux term at $\mathscr I^-(v_0)$ equals the flux terms across the remaining boundaries of $\mathscr O$. Using the Einstein equations, the flux term computed on the apparent horizon can be written as the derivative along the outgoing light-rays, of one quarter of the apparent horizon area. The flux conservation law then connects the variation in the horizon area to the variation of the relative entropy. Our main result is stated in \eqref{result}: we find that, to a variation of the relative entropy of the quantum matter, we can associate a variation of one-fourth of the apparent horizon area, plus contributions of the Kodama flux radiated toward future infinity, which can be set to zero for particular solutions of the Klein-Gordon equation. It is therefore natural to interpret the area as the entropy of the black hole.

The procedure outlined above is similar in spirit to the computation found in \cite{HollandsIshibashi19}. The key difference is in the choice of the initial data surface: we give initial data on the surface $\mathscr C$, while Hollands and Ishibashi gave initial data on the event horizon and at future null and time-like infinity. Our choice is more natural from a physical point of view, since we give initial data in the distant past and we observe the field evolution. On the other hand, this allows us to interpret the ground state as the vacuum state for modes coming from infinity. Finally, we do not need to make assumptions on the decaying properties of the field at time-like infinity, but we only need the asymptotic behaviour of the field at null infinity, which is uniquely determined by the assumption of asymptotic flatness. Non-trivial results on the decay properties of the scalar field at time-like infinity in a Schwarzschild background have been studied in \cite{DMP11}, but we do not know of similiar results on dynamical backgrounds.

The paper is organised as follows. In section \ref{geometry}, we introduce the geometric background of a dynamical, spherically symmetric, asymptotically flat black hole, we define its apparent horizon as the outer, future-directed trapping horizon introduced by Hayward \cite{Hayward93}, and we discuss the properties of the Kodama vector field and of its associated conserved current. Using the Stokes' theorem, we write the conservation law as an equation for the flux of the Kodama current across the boundary of the region $\mathscr O$. In section \ref{sec:2}, we review the algebraic formulation of a free quantum field theory on curved spacetimes, and we introduce the basics of Tomita-Takesaki modular theory in order to define the relative entropy between two states of a quantum field. We shall see that from the abstract Araki formula \eqref{eq:araki-formula}, it is possible to compute the relative entropy between coherent states in a much more concrete fashion, making use of the symplectic structure of the classical theory, \eqref{eq:EE-foundamental-formula}. We show that the derivative along the outgoing light-rays of the relative entropy between coherent states equals the Kodama flux at past null infinity, \eqref{kodama-current-relative-entropy}. In section \ref{sec:result}, we conclude the computation showing that the Kodama flux term on the apparent horizon equals the derivative of the horizon area along the outgoing light-rays, \eqref{flux-AH}. Using the Kodama flux conservation law, we can relate the derivative of the relative entropy to the derivative of one-quarter of the horizon area. The result is given in \eqref{result}. In section \ref{sec:limits}, we make contact with the thermodynamic approach to black hole entropy proposed by Hayward in \cite{Hayward97}. In the conclusions, we review the analysis of the paper and we propose further directions of research.

We work in units such that $G = \hbar = k_B = 1 $.
\section{Geometric setup} 
\label{geometry}
\subsection{Background metric} \label{subsec:spacetime}

In this section, we describe the black hole background and its symmetries, and the construction of the Kodama conservation law. We follow \cite{Wald84} for the notation and conventions in General Relativity.

The spacetime of interest is a time-dependent, time orientable, 4-dimensional, spherically symmetric black hole $\mathscr M$, whose metric $g$ can always be written as a warped product manifold between the radial-temporal plane and the 2-spheres of symmetry, $\mathscr M = \mathscr M_2 \times S_2$. Following \cite{Abreu10}, the background metric can be written in the form
\begin{equation} \label{metric}
\dd s^2 = x_{ab} \dd x ^a \dd x^b = \gamma_{ij}  \dd x^i \dd x^j + r^2(x) \sigma_{\alpha \beta} \dd \theta^\alpha \dd \theta^\beta \ ,
\end{equation}
where $\sigma_{\alpha \beta} \dd x^\alpha \dd x^\beta = \dd \theta^2 + \sin \theta \dd \varphi^2$ is the angular metric of the spheres of symmetry, $\gamma_{ij}$ is the metric on the radial-temporal plane, and $r$ is a function of the planar coordinates. We use indices from the beginning of the Latin alphabet ($a,b,c,...$) for the coordinates on the whole spacetime, indices from the middle of the Latin alphabet ($i,j,k,...$) for the planar components, and indices from the Greek alphabet ($\alpha, \beta, \gamma,...$) for the angular components. While this form of the metric highlights the warped product structure of the metric, it is independent on the choice of planar or angular coordinates. Assuming asymptotic flatness, the conformal structure at infinity coincides with that of a Schwarzschild black hole, and we can identify its boundaries with the past and future null infinities $\mathscr I^\pm$, time-like infinities $i^\pm$, and the space-like infinity $i^0$. The assumption of asymptotic flatness is motivated by the fact that we are interested in the formation and evaporation process of a black hole resulting from a spherical collapse, sourced by some local distribution of matter.

In spherical symmetry, it is useful to introduce the Israel-Hawking/Misner-Sharp  quasi-local mass
\begin{equation} \label{misner-sharp-mass}
m = \frac{r}{2}(1 - \nabla^a r \nabla_a r) \ .
\end{equation}
The Misner-Sharp mass represents active gravitational mass, satisfying the various Newtonian, small-sphere, large-sphere, special-relativistic, and test particle limits \cite{Hayward94}. 

With this form of the metric, the Einstein tensor satisfying the Einstein field equations for the background stress-energy tensor, $G = 8 \pi T^{(0)}$, takes the form \cite{Abreu10}
\begin{align} \label{Einstein-tensor}
G_{ij} &= - \frac{2 \nabla_i \nabla_j r}{r} + \bigg( \frac{2 \nabla^2 r}{r} - \frac{2m}{r^3} \bigg) \gamma_{ij} \\
G_{i \alpha} &= 0 \\
G_{\alpha \beta} &= \bigg( - \frac{r^2}{2} R_{ij} \gamma^{ij} + r \nabla^2 r \bigg) \sigma_{\alpha \beta}
\end{align}
where $R_{ij}$ is the planar component of the Ricci tensor.

Following Hayward \cite{Hayward97}, a sphere located at fixed $r$ is said to be \textit{untrapped}, \textit{trapped}, or \textit{marginal} if $\nabla^a r$ is respectively spatial, temporal, or null.  A hypersurface foliated by marginal spheres is called a \textit{trapping horizon}, which can be outer, degenerate, or inner if $\nabla^2 r > 0$, $\nabla^2 r = 0$, or $\nabla^2 r < 0 $ respectively, where $\nabla^2 = g^{ab}\nabla_a \nabla_b$. If $\nabla^a r$ is future- (past-) directed, the trapping horizon is said to be future (past). We then define the boundary of a black hole to be an outer, future trapping horizon, and we shall call it an \textit{apparent horizon}. The apparent horizon coincides with the event horizon for stationary black holes, but for dynamical spacetimes, it provides a local notion of a black hole boundary, while the event horizon is by definition a non-local concept. The apparent horizon is space-like if the black hole is growing, time-like if it is evaporating, and null if it is stationary.

In order to make explicit computations, we shall choose to write the metric as
\begin{equation} \label{metric-coordinates}
\dd s^2 = - e^{2\gamma(v,r)} f(v,r) \dd v^2 + 2 e^{\gamma(v,r)} \dd v \dd r + r^2 \dd \Omega^2 \ .
\end{equation}
$\gamma(v,r)$ is an arbitrary function of the coordinates, while $f(v,r)$ can be written in terms of the Misner-Sharp mass as
\begin{equation}
f(v,r) = 1 - \frac{2m(v,r)}{r} \ .
\end{equation}
The assumption of asymptotic flatness implies that $\gamma \to 0$ and $f \to 1$ at infinity. In these coordinates, the apparent horizon is located at $f = 0 \Rightarrow 2m(v,r) = r $. 

Finally, we introduce the null vector field
\begin{equation} \label{l}
l = e^{-\gamma(v,r)} \partial_v + \frac{f(v,r)}{2} \partial_r \ ,
\end{equation}
which describe the outgoing radial light-rays on the background metric.

\subsection{Kodama miracle} \label{subsec:kodama-miracle}
In dynamical spacetimes, it is in general not possible to define a global Killing vector field. However, Kodama showed \cite{Kodama79} that in spherically symmetric spacetimes it is possible to substitute the Killing vector field with the vector
\begin{equation} \label{Kodama}
k^i = \epsilon^{ij} \nabla_j r \ ,
\end{equation}
where $\epsilon^{ij}$ denotes the volume form associated with the planar metric $\gamma_{ij}$. In coordinates, the Kodama vector is
\begin{equation} \label{Kodama-coordinates}
k = e^{-\gamma} \partial_v \ .
\end{equation} 
From the definition, one can check that
\begin{equation} \label{kodama-r-property}
k_a \nabla^a r = 0 \ .
\end{equation}
The Kodama vector is divergence-free, and its associated conserved charge is just the areal volume $V = \frac{4 \pi}{3} r^3$,
\begin{equation} \label{Kodama-volume}
V = \int_{\Sigma} k_a \dd \Sigma^a
\end{equation}
where $\dd \Sigma^a$ is the oriented surface element of a 3-hypersurface, $\dd \Sigma^a = n^a \abs{\det h}^{1/2} \ \dd^3 y$, with $n^a$ the unit normal vector to the hypersurface, $h$ the induced 3-metric, and $y^a$ the induced coordinates on the hypersurface.

Associated to the Kodama vector we can define another conserved current,

\begin{equation} \label{Kodama-current}
j_a = T_{ab} k^b \ ,
\end{equation}
where $T_{ab}$ is a conserved stress-energy tensor. If we consider the Kodama current associated to the background stress-energy tensor, $j^a_{(0)} = T_{(0)}^{ab} k_b$, the corresponding conserved charge is the Misner-Sharp mass \cite{Hayward97},
\begin{equation} \label{mass-kodama}
m(q) - m(p) = - \int_{\Sigma} j^{(0)}_a \dd \Sigma^a \ ,
\end{equation}
where the integral is performed from the sphere of radius $r(p)$ to the sphere of radius $r(q)$ on a hypersurface $\Sigma$.

From its properties, one can see that the Kodama vector can be used to define a preferred flow of time along its integral curves \cite{Abreu10}. The Kodama vector is proportional to the time-like Killing field at infinity, where the spacetime is stationary, and its conserved current is naturally interpreted as the energy of the system, in analogy with the usual Minkowski case in which energy is the Noether charge associated to time translations. Moreover, from its definition we see immediately that the Kodama vector becomes null on the apparent horizon, just as the Killing vector is null on the event horizon.

To prove equation \eqref{mass-kodama}, we compute the contraction of the planar Einstein tensor with the Kodama vector:
\begin{equation}
G_{ij} k^j = - \frac{2}{r} \nabla_i ( \nabla_j r  ) \ \epsilon^{jk} \nabla_k r +\bigg( \frac{2}{r} \nabla^2 r - \frac{2m}{r^3} \bigg ) k_i \ .
\end{equation}
However, in $(1+1)-$dimensions we have
\begin{equation}
\epsilon_{[jk} \nabla_{i]} r = 0 \ ,
\end{equation}
and since the radial-temporal covariant derivative of the 2-dimensional volume form vanishes, we can write
\begin{multline}
\nabla_j \nabla_i r \epsilon^{jk} \nabla_k r =
\nabla_j (\epsilon^{jk} \nabla_i r ) \nabla_k r = \\
- \nabla^k r \nabla^j ( \epsilon_{ki} \nabla_j r + \epsilon_{ij} \nabla_k r ) = 
k_i \nabla^2 r - \frac{1}{2} \epsilon_{ij} \nabla^j (\nabla_k r \nabla^k r) \ .
\end{multline}
Using the definition of the Misner-Sharp mass \eqref{misner-sharp-mass}, we find
\begin{equation}
G_{ij} k^j = - \frac{1}{r} \epsilon_{ij} \nabla^j ( \frac{2m}{r} ) - \frac{2m}{r^3} \epsilon_{ij} \nabla^j r = - \frac{2}{r^2} \epsilon_{ij} \nabla^j m \ .
\end{equation}
Using the Einstein equations for the background stress-energy tensor, and using the symmetries to extend the relation to the full spacetime, we obtain
\begin{equation} \label{kodama-mass-gradient}
j^{(0)}_a = - \epsilon_{ab} \frac{\nabla^b m}{4 \pi r^2} \ .
\end{equation} 
From the above relation, equation \eqref{mass-kodama} follows.

We can now introduce the model we discussed in the \href{sec:intro}{Introduction}. We consider the propagation of a scalar, massless field $\hat \phi$ on the background, and we perturb it with a classical wave $\phi$, with non-vanishing initial data on the part of null past infinity with $v \geq v_0$, and with stress-energy tensor $T$, which is a coherent perturbation of the ground state. More precisely, the stress-energy tensor can be expanded in a perturbative series in the perturbative parameter $\lambda$,
\begin{equation}
T_{total} = T(0) + \lambda^2 T + o(\lambda^2) \ .
\end{equation}

$T(0) = T^{(0)}$ coincides with the stress-energy tensor of the source of the dynamical black hole and of the scalar field ground state, while $T = \frac{1}{2}\eval{\dv[2]{\lambda} T_{total}}_{\lambda = 0}$ is the stress-energy tensor of the coherent wave. The stress-energy tensor for the scalar field is understood as the expectation value
\begin{equation}
T_{ab}^\Phi = \omega\big(\partial_a \Phi \partial_b \Phi + \frac{1}{2} g_{ab} \partial_c \Phi \partial^c \Phi \big) 
\end{equation}
over the quasifree state $\omega$. The classical wave acts as a perturbation of the field, so that $\Phi = \hat \phi + \lambda \phi $, where $\hat \phi$ denotes the quantum scalar field, and the perturbative expansion of the expectation value of a term quadratic in the fields gives
\begin{equation}
\begin{split}
\omega(\Phi(x)\Phi(y)) &= \omega \big (\hat \phi(x)\hat \phi(y) \big ) + \\
&+ \lambda \bigg ( \omega \big (\hat \phi(x)\phi(y) \big ) + \omega \big (\hat \phi(y)\phi(x) \big ) \bigg )  + \\
&+ \lambda^2 \omega \big (\phi(x)\phi(y) \big ) + o(\lambda^2) \ .
\end{split}
\end{equation}
However, since $\phi$ is a classical wave, we have $\omega(\hat \phi(x)\phi(y)) = \phi(y) \omega(\hat \phi(x))$, and since, on quasifree states, any product of an odd number of fields vanishes (see the definition \eqref{quasi-free}), there is no linear correction to the stress-energy tensor. The classical wave therefore acts as a second order correction to the stress-energy tensor. The Einstein equations are then satisfied order by order, that is, we can perturbatively expand the Einstein tensor,
\begin{equation}
G_{total} = G^{(0)} + \lambda^2 \delta^2 G + o(\lambda^2)
\end{equation}
and set $\delta^2 G = 8 \pi T$.

From now on, we shall set $\lambda = 1$ for notational simplicity.

Since the coherent wave preserves spherical symmetry, the general form of the metric satifying the Einstein field equations for the full tensor $T_{total}$ remains the same as in \eqref{metric-coordinates}: the difference is that the arbitrary functions $\gamma(v,r)$ and $m(v,r)$ (or, equivalently, $f(v,r) = 1 - 2m/r$) appearing in the background metric get corrected, due to the coherent wave, by a second order perturbation, which replace in the definitions of the geometrical quantities $\gamma \to \Gamma(v,r) = \gamma(v,r) + \delta^2 \gamma(v,r)$, $m \to M(v,r) = m(v,r) + \delta^2 m(v,r)$. The location of the apparent horizon is then given by
\begin{equation}
r_{\mathscr {AH}} = 2M(v,r) \ ,
\end{equation}
that is, it is translated by $r_{\mathscr {AH}} - r^{(0)}_{\mathscr{AH}} = 2 \delta^2 m$.

We then construct the Kodama current associated with the full stress-energy tensor of the scalar field plus the source matter, $J_{tot}^a = T_{total}^{ab} k_b$. Since equation \eqref{kodama-mass-gradient} assumes only the general form for the Einstein tensor \eqref{Einstein-tensor}, which holds in any spherically symmetric spacetime, we can reproduce the same computation we did for the background current, arriving at
\begin{equation} \label{mass-kodama-full}
J^{tot}_a = - \epsilon_{ab} \frac{\nabla^b M}{4 \pi r^2} \ .
\end{equation}
Therefore, we have, as in \eqref{mass-kodama}, $M(q) - M(p) = - \int_\Sigma J_a^{tot} \dd \Sigma^a$.

 We now consider the Kodama current associated to the scalar field stress-energy tensor only, $j^a = T_{ab} k^b$. Thanks to the Stokes' theorem \cite{Wald84}, the conservation law for the Kodama current can be converted into an equation for its flux across the boundaries of a region $\mathscr O$ outside the black hole. We define $\mathscr O$ as the region bounded by four hypersurfaces: two null hypersurfaces at $v = v_0$ respectively $v = v_1$, the hypersurface $r = 2M$, with $v \in [v_0, v_1]$ and the corresponding part of past null infinity. We then consider the limit in which $v_1 \to \infty$. The resulting region, shaded in grey in figure \ref{diagram}, is a deformation of a double cone, in which one of the four null boundaries is substituted by the apparent horizon. We stress on the fact that we consider the apparent horizon for the full dynamics, located using the total Misner-Sharp mass $M$, instead of the background mass $m$. 

 We denote with $\mathscr{AH}(v_0)$ (respectively  $\mathscr I^-(v_0)$)  the part of the perturbed apparent horizon (resp. past null infinity) with $v \geq v_0$, and with $\Sigma_{v_0}$ the null hypersurface at $v = v_0$. The remaining parts of the boundary are the space-like infinity $i^0$ and the null ($\mathscr I^+$) and time-like ($i^+$) future infinity. However, since the coherent perturbation is switched on at $v = v_0$, the flux on $\Sigma_{v_0}$ vanishes for causality, as well as the flux at space-like infinity, since the coherent wave is a spatially compact solution of the Klein-Gordon equation. Therefore, the flux is naturally composed by three terms:
\begin{equation} \label{Kodama-flux-equation}
\Phi_{\mathscr{AH}} + \Phi_{\mathscr I^+} = \Phi_{\mathscr I^-} \ ,
\end{equation}
where each term is in the form $\Phi_\Sigma = \int_\Sigma j_a \dd \Sigma^a$, with the convention such that the directed surface element $\dd \Sigma^a$ is always future-directed. We implicitly consider the flux at time-like infinity in the flux term on the apparent horizon, extending the integral to $v = + \infty$.

The horizon term will give the derivative along the outgoing light-rays of one-fourth of the horizon area, while the term at past infinity will be written as the derivative of the relative entropy between the classical wave and the ground state. In the next section we need to take a brief \textit{tour} on the algebraic formalism for the quantization of free fields on globally hyperbolic spacetimes, with an obligatory visit to the Tomita-Takesaki modular theory to define the relative entropy via the Araki formula. At the end of the next section, we shall show that the relative entropy for coherent states equals the Kodama flux at $\mathscr I^-(v_0)$.

\section{Relative entropy between coherent states} \label{sec:2}
\subsection{Quantisation of the free, scalar, massless field on globally hyperbolic spacetimes} \label{subsec:quantisation}

In this section we review the basics of the algebraic approach to the free quantum scalar field, and the results on the relative entropy between coherent states found in \cite{Longo19}, \cite{CasiniGrilloPontiello19}, and \cite{CiolliLongoRuzzi19}. We refer to \cite{AAQFT15} and \cite{HollandsWald14} for a review on the algebraic quantisation.

On a spherically symmetric, dynamical black hole spacetime $\mathscr M = \mathscr M_2 \times S_2$, equipped with the metric $g$ given in \eqref{metric}, we consider a solution of the Klein-Gordon (K-G) equation,
\begin{equation}
\nabla^2 \phi = 0 \ ,
\end{equation}
where $\nabla^2 = g^{ab} \nabla_a \nabla_b$ is the K-G operator for a massless field with minimal coupling.

We define a symplectic space $(\mathsf{Sol}, w)$ as the space of smooth solutions of the K-G equation with spatially compact support, equipped with the symplectic form
\begin{equation}
\sigma(\phi_1, \phi_2) = \int_{\mathscr C}( \phi_2 \nabla_\mu \phi_1 - \phi_1 \nabla_\mu \phi_2) n^\mu \ \dd \Sigma \ ,
\end{equation}
where $\mathscr C$ is a Cauchy hypersurface (a surface whose causal development covers $\mathscr M$, see \cite{HawkingEllis73}) and $n$ is the unit, future-directed normal vector to $\mathscr C$. For the surface $\mathscr C$, we choose the limit in which part of the surface is light-like and coincides with past null infinity, as shown in figure \ref{diagram}, and we give non-vanishing initial data for the coherent wave on $\mathscr I^-(v_0)$, i.e., the part of past null infinity with $v > v_0$. The resulting classical wave represents a perturbation coming from the distant past. In this limit, the symplectic form becomes
\begin{equation} \label{eq:symplectic-form}
\sigma(\phi_1, \phi_2) = \int_{\mathscr I^-(v_0)}( \tilde \phi_2 \nabla_\mu \tilde \phi_1 - \tilde \phi_1 \nabla_\mu \tilde \phi_2) n_{\mathscr I^-}^\mu \ \dd v \dd \Omega \ ,
\end{equation}
where $n_{\mathscr I^-}$ is the unit, future-directed vector normal to $\mathscr I^-$, and $\tilde \phi = r \phi$.   It has been shown in \cite{DMP11} that, in the case of a Schwarzschild background, this is the correct limit to past infinity of a space-like Cauchy hypersurface; since our spacetime approaches the Schwarzschild spacetime in the asymptotic past, we can assume the same formula for the symplectic form.

Since the K-G operator is a Green hyperbolic operator, it admits unique advanced and retarded fundamental solutions $E^\pm$, whose integral kernels are distributions satisfying the Green equation $\nabla^2 E^\pm = \delta$, and are such that, for any test function $f \in C^\infty_0(\mathscr M)$, $\supp(E^\pm f) \subseteq J^\pm(\supp(f))$. A solution of the Cauchy problem for the K-G equation, with spatially compact support and initial Cauchy data given on a Cauchy surface $\mathscr C$, is given (non uniquely) by $\phi_f = E f$, where $f \in C^\infty_0(\mathscr M)$ and $E = E^- - E^+$ is the retarded-minus-advanced operator. Using $E$, it is possible to give a formula for the symplectic form which is independent on the chosen Cauchy surface; in fact, one can show that $E(f,g) = \sigma(\phi_f, \phi_g)$ \cite{AAQFT15}.

The algebra of quantum observables is obtained as the unique $\mathcal C^*-$algebra $\mathcal A$ generated by the abstract symbol $W(f)$ and the identity element $\mathbf 1$, satisfying the Weyl relations
\begin{align} \label{weyl-algebra}
W(0) &= \mathbf 1 \\
W(f)^* &= W(-f) = W(f)^{-1} \\
W(f) W(g) &= e^{- \frac{i}{2}\sigma(\phi_f, \phi_g)} W(f+g)
\end{align}
for any $f, \ g \in C^\infty_0(\mathscr M, \mathbb R)$. 

Alternatively, one can construct the quantum theory from the smeared quantum field $\hat \phi(f)$, treated as the abstract generator of a $*$-algebra. The symbol $W(f)$ is formally related to the more familiar smeared quantum field $\hat \phi(f)$ via exponentiation,
\begin{equation}
W(f) = e^{i \hat \phi(f)} \ .
\end{equation}

To compute expectation values of observables, we introduce a state as a linear, positive, normalized functional $\omega : \mathcal A \to \mathbb C$. We will consider only quasifree states in Hadamard form. Hadamard states are those states which admits a finite stress-energy tensor \cite{HollandsWald14}. Quasifree states are determined by their action on the Weyl operator, which is
\begin{equation} \label{quasi-free}
\omega(W(f)) = e^{- \frac{1}{2} \omega_2(f,f)}
\end{equation}
$\omega_2(f,f)$ uniquely determines the state. In the context of the $*$-algebra generated by $\hat \phi$, $\omega_2(f_1, f_2)$ is the 2-point function, $\omega_2(f_1, f_2) = \omega(\hat \phi(f_1) \hat \phi(f_2))$.

It is now possible to recover the usual description of a QFT as a theory of linear operators acting on a symmetrised Fock space via the GNS reconstruction theorem. The reconstruction is given in terms of a triple $(\mathcal H_\omega, \pi_\omega, \ket{\Omega_\omega})$, where $\mathcal H_\omega$ is a Hilbert space, $\pi_\omega$ is a representation of the algebra on the closable operators over $\mathcal H_\omega$, and $\ket{\Omega_\omega}$ is a unit vector in $\mathcal H_\omega$ such that $\omega(A) = \mel{\Omega_\omega}{\pi_\omega(A)}{\Omega_\omega}$ for any $A \in \mathcal A$. If $\omega$ is a quasifree state, it is possible to give a Fock representation of the $\mathcal C^*-$algebra, such that $\ket{\Omega_\omega}$ is the vacuum vector of the Fock space.

The GNS reconstruction theorem shows that, once a state is chosen, the algebraic and Hilbert space formulations of QFTCS are equivalent, with the crucial difference that there can be infinitely many unitary inequivalent representations of the algebra of observables $\mathcal A$. In Minkowski spacetime, the choice of a Poincaré invariant vacuum uniquely determines a preferred representation, but in less symmetric spacetimes the choice of a vacuum is not unique; in these contexts, the algebraic approach lets one discuss the properties of observables without referring to a particular state.

We now need to choose a quasifree, Hadamard state, which best represents a ground state on a dynamical black hole background. For a Schwarzschild black hole, the Unruh state \cite{Unruh76} is the most physically sensible candidate for the vacuum \cite{Candelas80} in both the interior and the exterior of the black hole, since it approaches a Minkowski vacuum at past infinity, exhibits Hawking radiation at late times, and is regular on the horizon. In particular, it has been proved that it is of Hadamard form \cite{DMP11}. 

For our purposes, however, since we are not interested in the details of the black hole collapse near the horizon, we shall consider a class of states. In particular, we fix the form of the 2-point function at past infinity only, assuming that it is regular in the sense of Hadamard on the rest of $\mathscr C$. The 2-point function restricted to $\mathscr C$ then defines the quasifree state in $D(\mathscr C)$, the causal development of $\mathscr C$, via the K-G equation. Since our spacetime coincides with the Schwarzschild spacetime at past infinity, we choose the 2-point function of a Unruh state on $\mathscr I^-$, so that the state is a vacuum with respect to modes coming from past infinity. Denoting $\tilde \phi_f = \eval{r \phi_f}_{\mathscr I^-}$, and remembering that we have non-vanishing initial data on $\mathscr I^-(v_0)$ only, the 2-point function is \cite{Kay88, Sewell82}
\begin{equation} \label{2-ptfunction}
\eval{\omega_2(f, g)}_{\mathscr I^-} = \frac{1}{\pi} \int_{\mathscr I^-(v_0)} \frac{\tilde \phi_f(v_1) \tilde \phi_g(v_2)}{(v_2 - v_1 - i \epsilon)^2} \ \dd v_1 \dd v_2 \dd \Omega \ ,
\end{equation}
where $\dd \Omega= \sin \theta \dd \theta \dd \varphi$ is the volume form of a unit 2-sphere. The integral is computed only on the part of null past infinity in which we give initial data, $\mathscr I^-(v_0)$. The 2-point function \eqref{2-ptfunction} is given by the Unruh 2-point function on $\mathscr I^-$ of a Schwarzschild black hole. Our choice for a class of ground states, to which we shall refer simply as \textit{the} ground state, is then defined by the requirement that the 2-point function is
\begin{equation}
\omega_2(f,g) =\eval{\omega_2(f,g)}_{\mathscr I^-} +\eval{\omega_2(f,g)}_{\mathscr C \ \setminus \ \mathscr I^-} \ .
\end{equation}
$\eval{\omega_2(f,g)}_{\mathscr C \ \setminus \ \mathscr I^-}$ is the 2-point function on the remaining part of the Cauchy surface, and it depends on the restrictions of $\phi_f, \phi_g$ and their normal derivatives to $\mathscr C \setminus \mathscr I^-$; we do not require a specific form for this part of the 2-point function, but only that is of Hadamard form.

This ground state can now be represented in the GNS reconstruction as the vacuum of a symmetrised Fock space, $\ket \Omega$. The coherent wave $\phi$, propagating over the spacetime with initial data on $\mathscr I^-(v_0)$ is then a coherent perturbation of the ground state, which can be represented as $\ket \Phi = W(f) \ket \Omega$, where $\phi = Ef$. We denote the corresponding state functional and its GNS vector representative with with $\omega_\phi(\cdot) = \mel{\Phi}{ \ \cdot \ }{\Phi}$. 

    \subsection{Tomita-Takesaki modular theory}
In this section, we introduce the basics of the Tomita-Takesaki modular theory in order to discuss the Araki formula \eqref{eq:araki-formula} for the relative entropy \cite{Araki76}. The goal is to state formula \eqref{eq:EE-foundamental-formula} as a means to compute the relative entropy between coherent states. We refer to \cite{CasiniGrilloPontiello19} and \cite{CiolliLongoRuzzi19} for the recent results on the properties of relative entropy between coherent states,  while we refer to \cite{Haag67} for a classical review on the subject of modular theory. For a more physically-oriented introduction, we refer to \cite{Witten18}.

We start very general, and we slowly reduce to the case of coherent states for a QFT. We shall prove the properties of Tomita-Takesaki theory which are most useful for the characterization of the relative entropy of coherent states, with no attempt to be exhaustive.

Let $ \mathfrak U(\mathcal H) $ be a von Neumann algebra on a Hilbert space $ \mathcal H $, and let $\ket \Omega \in \mathcal H$ be a cyclic and separating vector, which means that, i) given a vector $\ket \Omega \in \mathcal H$ and an operator $\pi(A)$ over $\mathcal H$, $ \pi(A) \ket \Omega$ is dense in $ \mathcal H $, and ii) $\pi(A) \ket \Omega = 0 \Rightarrow \pi(A) = 0 \ \forall \ A \in \mathcal A$. Then, there exists a unique antilinear operator $ \mathcal S_\Omega$, called the \textit{modular involution} or \textit{Tomita operator}, such that
\begin{equation} \label{def:tomita-operator}
   \mathcal S_{\Omega} A \ket{\Omega} = A^* \ket{\Omega} \ .
\end{equation}
From the definition, it is clear that $\mathcal S^2_\Omega = 1 $, and therefore it is invertible. Moreover, $ \mathcal S_\Omega \ket \Omega = \ket \Omega $.

An invertible, closed operator always admits a unique polar decomposition, $\mathcal S_\Omega = J_\Omega \Delta_\Omega^{1/2}$, where the \textit{modular conjugation} $J_\Omega$ is an anti-linear, unitary operator and the \textit{modular operator} $\Delta_\Omega$ is self-adjoint and non-negative. (The Tomita operator as defined in \eqref{def:tomita-operator} is not closed, but is closable. We denote the closure of $\mathcal S_\Omega$ with the same symbol.)

As $\mathcal S_\Omega \ket \Omega = \mathcal S_\Omega^* \ket \Omega = \ket \Omega $, it is immediate to see that $ \Delta_\Omega \ket \Omega = \ket \Omega $ and therefore $ J_\Omega \ket \Omega = \ket \Omega $. Moreover, from $\mathcal S^2_\Omega = 1$, we have $J^2_\Omega = 1$ and $J_\Omega \Delta_\Omega^{is} J_\Omega = \Delta_\Omega^{is} $, with $s \in \mathbb R$.

The \textit{modular Hamiltonian} is then defined as
\begin{equation}
    K_\Omega = - \log \Delta_\Omega \ .
\end{equation}
$K_\Omega$ is a self-adjoint operator with generally unbounded spectrum, which is well-defined since $\Delta_\Omega$ is non-negative. By the Stone theorem, $K_\Omega$ defines a 1-parameter group of unitary operators on the von Neumann algebra,
\begin{equation} \label{alpha}
\alpha_s(A) = e^{-iK_\Omega s}A e^{iK_\Omega s} = \Delta_\Omega^{is} A \Delta_\Omega^{-is} \ .
\end{equation}
The one-parameter group of unitaries $ \Delta^{is} $ is called \textit{modular flow}.

We can as well define the Tomita operator for the commutant of $ \mathfrak U(\mathcal H) $, that is, the set of bounded operators $\mathfrak U'(\mathcal H) = \{ U' \ | \ [U,U'] = 0 \  \forall \ U \in \mathfrak U(\mathcal H) \} $, and perform the same polar decomposition with the same properties as above. The relations between the von Neumann algebra, the modular flow, and its commutant are then given by the relations \cite{Takesaki70}
\begin{align}
    & J_\Omega \mathfrak U(\mathcal H) J_\Omega = \mathfrak U(\mathcal H)' \label{commutant}\\
    & \Delta_{\Omega}^{is} \mathfrak U(\mathcal H) \Delta_{\Omega}^{-is} = \mathfrak U(\mathcal H) \quad \ , \  \quad\Delta_{\Omega}^{is} \mathfrak U(\mathcal H)' \Delta_{\Omega}^{-is} = \mathfrak U(\mathcal H)' \label{modular-flow}
\end{align} 
The first property says that the modular conjugation maps the algebra into its commutant. The second one states that the modular flow defines an automorphism of the algebra. 

The action of the modular operator on the observables is in general very difficult to evaluate, but for free theories on flat spacetimes the \textit{Bisognano-Wichmann Theorem} \cite{Bisognano75} links the modular flow to a geometric action.

In Minkowski $\mathbb M = \mathbb R^{1,3}$, we consider the region $\mathscr W = \{ \mathbf x \in \mathbb R^{1,3} \ | \ x > \abs{t}\} $, called \textit{Rindler wedge}. This is the domain of dependence of the surface $\Sigma = \{ \mathbf x \in \mathbb R^{1,3} \ | \ t=0, \ x > 0 \} $. Consider the vacuum vector $\ket \Omega$ for the theory defined on the whole Minkowski spacetime $\mathfrak U(\mathscr M) $. By the Reeh-Schlieder theorem \cite{Witten18}, this is a cyclic and separating vector for the algebra of observables restricted to the Rindler wedge $\mathfrak U(\mathscr W)$. Then, the Bisognano-Wichmann theorem states that \cite{Bisognano75}
\begin{equation}\label{th:bisognano-wichmann-theorem}
J_\Omega = \Theta U(R_1(\pi)) \quad \Delta_\Omega = e^{-2 \pi K_1} \ ,
\end{equation}
where $\Theta$ is the CPT operator, $U(R_1(\pi)) $ is the unitary operator representing a space rotation of $ \pi $ degrees around the $x$ axis and $K_1$ is the generator of the one-parameter group of boosts in the plane $(t,x)$. The theorem admits a generalisation \cite{Brunetti02} to the algebras defined on a spacetime with a group of symmetries, in which a "wedge region" can be defined. Such a generalisation is what interests us, in order to compute the relative entropy on the half-infinite part of past null infinity with $v > v_0$.
    
    \subsection{Relative entropy}
Now, we consider the induced von Neumann algebra $\mathfrak U(\mathcal A)$ as the GNS representation of a $\mathcal C^*-$algebra $\mathcal A$, defined by a state functional $\omega$, and the cyclic and separating vector $\ket \Omega$ representative of  $\omega$. If we consider the so-called \textit{natural cone}, that is, the set of vectors
\begin{equation}
\mathcal P_\Omega = \overline{\{ A j_\Omega(A) \ket{\Omega} | \ A \in \mathfrak U \}} \ ,
\end{equation}
where the bar means the closure and $j_\Omega(A) = J_\Omega A J_\Omega$,  another state $\omega_\Phi$ has a unique representative vector $\ket \Phi$ in $\mathcal P_\Omega$, represented in the usual way, $\omega_\Phi(A) = \ev{A}{\Phi} $. The one-to-one correspondence between vectors and states comes at hand when we want to discuss the properties of states under some automorphism, and in turn does not introduce technical difficulties.

Now, given two cyclic and separating vectors, $\ket{\Omega} \in \mathcal H, \ \ket{\Phi} \in \mathcal P_\Omega$, we can generalise the construction of the entropy to the notion of relative entropy. We define the \textit{relative Tomita operator} (or \textit{relative modular involution})  as
\begin{equation} \label{eq:relative-tomita-operator}
    \mathcal S_{\Omega, \Phi} A \ket{\Phi} = A^* \ket{\Omega} \ ,
\end{equation}
which admits the unique polar decomposition $\mathcal S_{\Omega, \Phi} = J_{\Omega, \Phi} \Delta_{\Omega, \Phi}^{1/2}$. Since $\ket \Phi$ is in the natural cone, one can show that $ J_{\Omega, \Phi} = J_\Omega$ \cite{Witten18}. The relative modular Hamiltonian is defined as
\begin{equation}
    K_{\Omega, \Phi} = - \log \Delta_{\Omega, \Phi} \ .
\end{equation}
The Araki formula then defines the relative entropy:
\begin{equation} \label{eq:araki-formula}
    S(\omega_\Omega | \omega_\Phi) = \ev{\log \Delta_{\Omega, \Phi}}{\Omega} = - \ev{K_{\Omega, \Phi}}{\Omega} \ .
\end{equation}
In the case of Quantum Mechanics, where states are realised as density matrices, $\omega(A) = \Tr \rho_\Omega A $, the relative entropy takes the familiar form $S(\Omega|\Phi) = - \Tr \rho_\Omega (\log\rho_\Phi - \log \rho_\Omega) $ \cite{Araki76}, the vacuum-subtracted entanglement entropy we mentioned in the  \href{sec:intro}{Introduction}.

\subsection{Relative entropy for coherent states} \label{ssec:EE-coherent-states}
Now, following \cite{CasiniGrilloPontiello19}, we take a cyclic and separating vector $\ket \Omega$ and a vector $\ket \Phi \in \mathcal P_\Omega$, $\ket \Phi = j_\Omega(U) U \ket \Omega$, for some unitary operator $U \in \mathfrak U(\mathcal H)$. The operator $j_\Omega(U) U$ is still a unitary operator, and the corresponding state functional is $\omega_\Phi(A) = \braket{\Omega U}{A U \Omega} = \omega(U^*AU)$.

From property \eqref{commutant}, acting with $U \mathcal S_{\Omega} j_\Omega(U^*) A$ on $\ket \Phi$ for some operator $A$, we can see that
\begin{equation}
( U \mathcal S_{\Omega} j_{\Omega}(U^*) ) \ A \ket{\Phi} =
         U \mathcal S_{\Omega} (AU) \ket{\Omega} = A^* \ket{\Omega} = \mathcal S_{\Omega, \Phi} A \ket{\Phi} \ .
\end{equation}
Therefore, the relative Tomita operator between $\ket \Omega$ and $\ket \Phi$ can be computed using the Tomita operator only:

\begin{equation}\label{relative-tomita-and-tomita-relation}
       \mathcal S_{\Omega, \Phi} = U \mathcal S_{\Omega} j_{\Omega}(U^*)  \ .
\end{equation}
By polar decomposition, one can see that
\begin{equation}
\Delta^{1/2}_{\Omega, \Phi} = j_{\Omega}(U) \Delta^{1/2}_{\Omega} j_{\Omega}(U^*) \ \ K_{\Omega, \Phi} = j_\Omega(U) K_\Omega j_\Omega(U^*) \ .
\end{equation}

From the above relation, by direct computation we can see that the relative entropy between $\ket \Omega$ and $\ket \Phi$ can be computed from the modular operator only:
\begin{multline}
    S(\omega_{\Omega} | \omega_{\Phi} ) = \ev{\log\Delta_{{\Omega}, U \Omega}}{\Omega} = \\ 
    = 2  \ev{\log\Delta_{\Omega,\Phi}^{1/2}}{\Omega} = 2 \ev{\log(j_{\Omega}(U) \Delta^{1/2}_{\Omega} j_{\Omega}(U^*))}{\Omega} \ .
\end{multline}
Since $  \log (j_{\Omega}(U)\Delta_{\Omega}j_{\Omega}(U^*) ) =j_{\Omega}(U) \log \Delta_{\Omega} j_{\Omega}(U^*) $, using the invariance of $\ket \Omega$ under $J_\Omega$, and the fact that $J_\Omega \Delta_\Omega J_\Omega = \Delta_\Omega^{-1}$, it is straightforward to show that:
\begin{equation} \label{eq:relative-entropy-modular-hamiltonian}
S(\omega_\Omega |\omega_{U\Omega} ) = \ev{K_\Omega}{U^*\Omega} = i \eval{\dv{t}\ev{e^{-iKt}}{U^* \Omega}}_{t=0} \ .
\end{equation}

We now specialise to the case in which $\omega_\Omega$ is the ground state functional, which will denote simply by $\omega$, with vector representative $\ket \Omega$, and $\omega_{U \Omega}$ is a coherent state with $U = W(f)$, for some test function $f$, which will be denoted by $\omega_f(A) = \ev{A}{\Phi} = \omega(W^*(f) A W(f))$. We use $f$ as a label to remember that $\omega_f$ is a coherent state, and can be considered a classical perturbation $\phi_f= Ef$ of the ground state. The unique representative vector of $\omega_f$ in the natural cone $\mathcal P_\Omega$ is $\ket \Phi = j_\Omega(W(f))W(f) \ket \Omega$. 

The entropy for coherent states is computed using the classical structure only, that is, from the symplectic form of $\mathsf{Sol}$. In fact, substituting the definitions of coherent state in \eqref{eq:relative-entropy-modular-hamiltonian}, i.e., choosing $U = W(f)$, we have
\begin{equation}
S(\omega | \omega_f) = i \eval{ \dv{t} \ev{W(f)^* \Delta^{it}W(f)}{\Omega}}_{t=0} \ .
\end{equation}
Using the invariance of the vacuum with respect to $\Delta_\Omega$, it is possible to rewrite this expression as the action of the automorphism $\alpha_t$ introduced in \eqref{alpha} over the Weyl operator:
\begin{multline}
i \eval{\dv{t} \ev{W^*(f)\Delta^{it} W(f) \Delta^{-it}}{\Omega}}_{t=0} = \\
= i \eval{\dv{t} \ev{W^*(f) \alpha_tW(f)}{\Omega}}_{t=0} \ .
\end{multline}
Using the Weyl relations, we have
\begin{equation}
W^*(f)\alpha_t\big (W(f) \big ) = e^{-\frac{i}{2} \sigma(-\phi_f, \alpha_t(\phi_f))}W(\alpha_t(f)-f)
\end{equation}
and
\begin{equation}
 \ev{W(f)}{\Omega} = e^{-\frac{1}{2}\omega_2(f,f)} \ ,
\end{equation}
from which we can write the expectation value as
\begin{multline} \label{expval-intermediate}
\ev{W^*(f)W(\alpha_t(f))}{\Omega} = \\ 
= e^{-\frac{i}{2} \sigma (-\phi_f, \alpha_t(\phi_f))}\ev{W(\alpha_t(f)-f)}{\Omega} = \\
\exp{-\frac{i \sigma (-\phi_f, \alpha_t(\phi_f))}{2} - \frac{1}{2}\omega_2(\alpha_t(f) - f, \alpha_t(f) - f) } \ .
\end{multline}

Now, we have
\begin{equation}
\eval{\dv{t} \omega_2(\alpha_t(f) - f, \alpha_t(f) - f)}_{t= 0} = 0 \ ,
\end{equation}
since $\eval{\alpha_t(f)}_{t = 0} = f$. Taking the derivative with respect to $ t $ of \eqref{expval-intermediate}, evaluated at $t = 0$, we notice that the exponential factor vanishes for $t = 0$, obtaining
\begin{equation}
i \dv{t} \eval{\ev{W^*(f)W(\alpha_t(f))}{\Omega}}_{t=0} = \frac{1}{2} \sigma \bigg( \eval{\dv{\alpha_t(\phi_f)}{t}}_{t=0}, \phi_f \bigg)  \ .
\end{equation}
This conclude the derivation of the main result of this section, which can be found (each in a slightly different derivation) in \cite{HollandsIshibashi19, Longo19, CasiniGrilloPontiello19}: the relative entropy between a coherent state and the vacuum is given by the derivative of the symplectic form of the associated classical solution:
\begin{equation} \label{eq:EE-foundamental-formula}
    S(\omega | \omega_f) = \frac{1}{2} \sigma(\eval{\dv{t} \alpha_t(\phi_f)}_{t=0}, \phi_f) \ .
\end{equation}

We can now use the above formula to compute the relative entropy between the ground state defined by \eqref{2-ptfunction} and a coherent perturbation with the symplectic form given in \eqref{eq:symplectic-form}, on the dynamical black hole background. We first notice that, from the expressions \eqref{2-ptfunction} and \eqref{eq:symplectic-form}, it is clear that these equations actually defines an algebra on the half-infinite part of past null infinity $\mathscr I^-(v_0)$, which coincides with the restriction of the quantum theory on the whole spacetime, $\mathcal A(\mathscr M)$, to the past null infinity, $\mathcal A(\mathscr I^-(v_0))$. On such a spacetime, we can apply the generalisation of the Bisognano-Wichmann theorem by Brunetti, Guido, and Longo \cite{Brunetti02}, which states that the modular action $\alpha_t(\phi(f))$ acts as a boost in the radial-temporal plane: in our coordinates, the modular action becomes
\begin{equation}
\alpha_t(\phi(v, \theta, \varphi)) = \phi(v_0 + e^{-2 \pi t}(v - v_0), \theta,  \varphi) \ .
\end{equation}
Therefore, the relative entropy \eqref{eq:EE-foundamental-formula} in the region $v > v_0$ becomes
\begin{multline} \label{relative-entropy-coherent-states}
S(\omega | \omega_\phi) = \\
=  - \pi \int_{\mathscr I^-(v_0)} \big [ \tilde \phi \partial_a [(v - v_0) \partial_v \tilde \phi ] - (v - v_0) \partial_v \tilde  \phi \partial_a \tilde \phi \big ] n^a_{\mathscr I^-} \dd v \dd \Omega \ .
\end{multline}

We can integrate by parts the first term in the integral, obtaining
\begin{equation} \label{RE-past-infinity}
S(\omega | \omega_\phi) = 2 \pi \int_{\mathscr I^-(v_0)} (v - v_0) \partial_a \tilde \phi \partial_b \tilde \phi \ n_{\mathscr I^-}^a k^a \ \dd v \dd \Omega \ ,
\end{equation}  
where we used $k^a_{\mathscr I^-} = \partial_v$.

Due to the decaying properties of $\phi$ at past null infinity on a Schwarzschild background \cite{DMP11}, this entropy formula remains finite. 

In order to make the connection with the conservation law for the Kodama flux, \eqref{Kodama-flux-equation}, we rewrite the above expression in terms of the stress-energy tensor of the scalar field at infinity, 
\begin{equation}
T_{ab} = \partial_a \tilde \phi \partial_b \tilde \phi - \frac{1}{2} g_{ab} \partial_i \tilde \phi \partial^i \tilde \phi \ .
\end{equation}
Since the metric is anti-diagonal at past infinity, and the vectors $k$ and $n_{\mathscr I^-}$ are parallel, we have $g_{ab} k^a n^b_{\mathscr I^-} = 0$. Therefore we can substitute the stress-energy tensor in the relative entropy formula, finding
\begin{equation}
S(\omega | \omega_\phi) = 2 \pi \int_{\mathscr I^-(v_0)} (v - v_0) T_{ab} \ n_{\mathscr I^-}^a k^a \ \dd v \dd \Omega \ .
\end{equation}

We now consider what happens if we rigidly translate the boundary of the region in which the field propagates by a finite amount, $v_0 \to v_0 + \tau$. The derivative of the relative entropy with respect to this rigid translation is
\begin{equation} \label{derivative-RE-kodama}
- \frac{1}{2 \pi}\dv{v_0}S(\omega | \omega_\phi) = \int_{\mathscr I^-(v_0)} T_{ab} k^b n^a_{\mathscr I^-} \ \dd v \dd \Omega \ .
\end{equation}
Such a derivative can be interpreted as the derivative with respect to the instant $v$ in which we switched on the perturbation, evaluated for $v = v_0$.

 Finally, we notice that the right-hand side of the above equation equals (by definition) the Kodama flux across past null infinity of equation \eqref{Kodama-flux-equation}, while the derivative with respect to $v_0$ can be understood as the derivative along the outgoing light-rays, evaluated at $v = v_0$:
\begin{equation} \label{kodama-current-relative-entropy}
- \frac{1}{2 \pi} l(S(\omega | \omega_\phi))(v_0) = \Phi_{\mathscr I^-} \ .
\end{equation}
\section{Variation of generalised entropy} \label{sec:result}

We now return to the original task of computing the various terms in the conservation law for the Kodama flux, \eqref{Kodama-flux-equation}. Of the three terms, we have just seen that $- \frac{1}{2 \pi}l(S(\omega | \omega_\phi))(v_0) = \Phi_{\mathscr I^-}$. On the other hand, on the apparent horizon of the perturbed spacetime, we consider the flux term
\begin{equation}
\Phi_{\mathscr{AH}} = \int_{\mathscr{AH}(v_0)} j_a \dd \Sigma^a_{\mathscr{AH}} \ ,
\end{equation}
where the integral is extended over the part of apparent horizon with $v > v_0 $. 

Using \eqref{mass-kodama}, it is immediate to see that the flux term equals the mass contribution to the black hole by the scalar field,
\begin{equation}
\Phi_{\mathscr{AH}} =  \delta^2 m(v_0, r_{\mathscr{AH}}) - \delta^2 m_{i^+} \ ,
\end{equation}
where $\delta^2 m_{i^+} = \delta^2 m(v = \infty, r_{\mathscr{AH}})$.

To see this more explicitly, consider the Kodama currents constructed from the full stress-energy tensor $J^{tot}_a = T^{tot}_{ab}k^b$ and from the background matter, $j^{(0)}_a = T^{(0)}_{ab} k^b$. Then we have $j  = J^{tot} - j^{(0)}$, and again using \eqref{mass-kodama} and \eqref{mass-kodama-full}, the integration of the two Kodama currents on the apparent horizon gives the desired result, since by definition $M - m = \delta^2 m$.

On the other hand, we can compute the derivative of $A = 4 \pi r^2$ along the \textit{background} outgoing light-rays on the apparent horizon, $l(A)_{\mathscr{AH}}$. As we said, the apparent horizon is not at $r = 2m$, but gets translated to $r = 2M(v,r)$. This implies that $\eval{f}_{\mathscr{AH}} = \frac{1}{2M}(2M - 2m)$, and so
\begin{equation}
\eval{l(A)}_{r = 2M} = \frac{2M - 2m }{4M} \eval{\partial_r A}_{r = 2 M} = 8 \pi (M - m) = 8 \pi \delta^2 m  \ .
\end{equation}
Evaluating the above equation for $v = v_0$, we find that
\begin{equation} \label{flux-AH}
\Phi_{\mathscr{AH}} = \frac{1}{8 \pi} l(A)_{\mathscr{AH}}(v_0) - \delta^2 m_{i^+} \ .
\end{equation}

We can now conclude our computation of the Kodama flux. In the flux conservation law \eqref{Kodama-flux-equation}, we have shown that the apparent horizon term can be rewritten as a variation of the area of the horizon, giving \eqref{flux-AH}; on the other hand, the past infinity term has been rewritten as a variation of the relative entropy of the perturbation, via the Araki formula for coherent states \eqref{eq:EE-foundamental-formula}, obtaining \eqref{kodama-current-relative-entropy}. Substituting equations \eqref{flux-AH} and \eqref{kodama-current-relative-entropy} in the flux conservation law \eqref{Kodama-flux-equation}, we get
\begin{equation} \label{result}
l( S(\omega | \omega_\phi))_{\mathscr I^-}(v_0) + \frac{1}{4} l(A)_{\mathscr{AH}}(v_0)= 2 \pi \Phi_{\mathscr I^+} + \delta^2 m_{i^+} \ .
\end{equation}

Equation \eqref{result} generalises the result by Hollands and Ishibashi \cite{HollandsIshibashi19}, from the event horizon of a static black hole to the case of apparent horizons of dynamical, spherically symmetric black holes.

To make the result clearer, we can consider a coherent wave such that $\phi \to 0$ for large $v$, and the right-hand side of the equation above vanishes automatically. It is then straightforward to integrate the left-hand side along the geodesic congruence tangent to $l$. Choosing the surface $v = v_0$ as the surface on which the integral parameter $\tau$ of $l$ vanishes, and integrating between $0$ and $+ \infty$, that is, from the surface $v = v_0$ to $\mathscr I^+$, we find
\begin{equation}
\Delta S_{gen} = c(r, \theta, \varphi) \ ,
\end{equation}
where it is natural to introduce a generalised entropy $S_{gen} = S(\omega | \omega_\phi) + \frac{1}{4} A$, and $c(r, \theta, \varphi)$ is an arbitrary function of integration.

\section{Thermodynamic Interpretation}  \label{sec:limits}

To conclude, we want to make contact with the first law of dynamical black holes found by Hayward \cite{Hayward97}. In his paper, he introduced two invariant quantities, the scalar
\begin{equation}
w = - \frac{1}{2} \gamma_{ij} T^{ij}
\end{equation}
and the vector
\begin{equation}
\psi = T_{ab} \nabla^b r + w k_a \ .
\end{equation} 
$w$ is an energy density, while $\psi$ plays the role of a localized Bondi energy flux. He then showed that, along any vector $t$ tangent to the apparent horizon, one has
\begin{equation} \label{hayward-law}
- \Phi_{\mathscr {AH}(v_0)} = \int_{\mathscr{AH}} \dd m = \int_{\mathscr{AH}} \big ( \frac{\kappa_{\mathscr {AH}}}{8 \pi} \nabla_a A + w \nabla_a V \big ) t^a \dd^3 y \ ,
\end{equation}
where $\kappa_{\mathscr{AH}}$ is the surface gravity on the apparent horizon associated to the Kodama vector,
\begin{equation}
k^a \nabla_{[a} k_{b]} = \kappa k_b \ .
\end{equation}
Equation \eqref{hayward-law} resembles (and, in fact, coincides with, see \cite{Hayward97} for the details) the first law of thermodynamics, $\dd E = T \dd S + p \dd V$. Now, we want to show that the term at past infinity can also be interpreted as a thermodynamic contribution to the system.

In fact, if we introduce a new coordinate $V = e^{\kappa (v - v_0)} $, a boost acts in this case as
\begin{equation}
V \to e^{\kappa e^{-2 \pi t} (v-v_0)} = V^{e^{-2 \pi t}} \ .
\end{equation}
The modular action on the scalar field restricted to $\mathscr I^-$ therefore is
\begin{equation}
\eval{\dv{t} \alpha_t(\phi(V, \theta, \varphi))}_{t = 0} = - 2 \pi V \partial_V \phi(V, \theta, \varphi) \ .
\end{equation}
On the other hand, the Kodama vector at $\mathscr I^-$ in this new coordinate is $k = \pdv{V}{v} \partial_V = \kappa V \partial_V$, and therefore, computing \eqref{eq:EE-foundamental-formula} in these coordinates, following the same passages we did in section \ref{ssec:EE-coherent-states},  we arrive at
\begin{equation}
S_{v_0}(\omega| \omega_\phi) = - 2 \pi \int_{\mathscr I^-(v_0)} \frac{1}{\kappa_{\mathscr I^-}} T_{ab} k^a \dd \Sigma^b
\end{equation}
Since $\kappa_{\mathscr I^-}$ is a constant at infinity, we get
\begin{equation}
\frac{1}{2 \pi}\kappa_{\mathscr I^-} S_{v_0}(\omega|\omega_\phi) = - \Phi_{\mathscr I^-} \ .
\end{equation}
We see then that the relative entropy has a clear thermodynamic interpretation as a true entropy contribution from the scalar field, where the temperature is given by $T = \frac{\kappa}{2 \pi}$, which is analogous to the Hawking temperature for a static black hole. The Kodama flux is the energy contribution of the scalar field to the black hole. The minus sign comes because we choose as positive the flux exiting the black hole, while this is the energy contribution of the environment on the system. A detailed discussion on the role of $\kappa/2 \pi$ as a physical measure of the temperature of a dynamical, spherically symmetric black holes, in terms of the temperature of the states of a scalar field near the apparent horizon, can be found in \cite{KPV21}.

\section{Conclusions and outlook}
In this paper, we showed that, using the variation of the relative entropy between coherent states of a quantum scalar field, it is possible to assign to a dynamical black hole with spherical symmetry an entropy proportional to the area of the apparent horizon. The result reduces to the Bekenstein-Hawking formula for stationary black holes, and it agrees with the result obtained by Hayward \cite{Hayward97} by means of rewriting the first law of thermodynamics in a general-relativistic setting. Our computation generalises the analysis for Schwarzschild black holes made in \cite{HollandsIshibashi19}, with the difference that we give initial data in the asymptotic past, and it shows that a direct computation of the variation of the entropy of the quantum matter give rise to a natural notion of entropy for black holes, without introducing any new assumption on the black hole physics. Moreover, we see that, at least for the case of coherent perturbations, the relative entropy approach solves two of the problems emerged using the entanglement entropy approach \cite{Bombelli86}, \cite{Solodukhin11}, that are, the divergence in the continuum limit and the dependence on the number of fields. As proved in \cite{HollandsIshibashi19}, the relative entropy formula \eqref{relative-entropy-coherent-states} is finite. On the other hand, if we had more than one field, we could simply sum each contribution to get the relative entropy for the matter fields, while the sums of the stress-energy tensors would give the total variation of the area.

The two key ingredients for the computation are asymptotic flatness and spherical symmetry. Asymptotic flatness lets us use the results for the relative entropy between coherent states obtained in Minkowski \cite{Araki76}, \cite{Longo19} and Schwarzschild \cite{HollandsIshibashi19} spacetime, in the curved setting of a dynamical black hole spacetime. Thanks to asymptotic flatness, we can study the structure of the spacetime at infinity using the conformal treatment of Penrose diagrams \cite{Penrose64}. Moreover, our dynamical background asymptotically coincides with a Schwarzschild background, and we can use the decaying properties of the scalar field towards past null infinity studied in \cite{DMP11}, in order to write the symplectic form \eqref{eq:symplectic-form} as a finite integral.  The assumption of spherical symmetry, on the other hand, introduces the geometric conservation law which is central to our derivation, to connect the variation of entropy at infinity to the variation of the horizon area.

The assumption of asymptotic flatness seems well motivated in the context of black hole physics, in which one is interested in the physics of a black hole perturbed by some distribution of matter, but which is otherwise isolated from distant effects. It would be interesting to study what happens in physical settings in which asymptotic flatness does not hold, as would be the case for a black hole spacetime with non-vanishing cosmological constant.

On the other hand, observed, astrophysical black hole are axisymmetric, so that the assumption of spherical symmetry should be abandoned in favor of more realistic models. It seems likely that a derivation for stationary, axisymmetric black holes could follow the same lines of the computation done here and in \cite{HollandsIshibashi19}, due to the presence of a Killing vector field. A derivation for the case of dynamical, axisymmetric black holes requires further investigation, since the Kodama conservation law is not at disposal. On the geometric side, it might be interesting to apply the same methods for the case of cosmological event and apparent horizons, to see if the relative entropy could explain the same area law for the entropy of cosmological horizons \cite{Gibbons77}.

On the quantum side, it would be interesting to study the relative entropy for different quantum fields. Although, for simplicity, we considered the massless scalar field only, the computation of formula \eqref{relative-entropy-coherent-states} relies only on the Weyl relations and the asymptotic behaviour of the fields, and so we can expect to generalises the computations for boson fields of higher spin. It would be interesting to attempt the same computation for coherent states of fermions.

The second generalisation on the matter side would be to consider more general states. The difficulty in this case would be to generalise the computation of the relative modular operator to different classes of states; in, e.g., \cite{Drago17}, the relative entropy has been computed for certain classes of thermal states in flat spacetime, and one direction of research could be to generalise their results to curved spacetimes.

Although we are not able to justify the microscopic origin of the black hole entropy, in \cite{HollandsIshibashi19} it has been shown that the same computation holds for gravitational, coherent perturbations in a stationary background, assuming that the decaying properties of the scalar field holds for gravitational perturbations, too. It would be interesting to see if their result generalises to higher perturbative orders, or in a non-perturbative treatment of quantum gravity.

As a final note, we remark that our computation of the relative entropy for coherent states relies on the restriction of the theory to the conformal boundary of the spacetime. In particular, we use the restriction to compute the modular flow for the algebra on conformal past infinity $\mathcal A(\mathscr I^-(v_0))$, in equation \eqref{relative-entropy-coherent-states}, seen as a restriction of the algebra on the causal development of $\mathscr C$. It may be interesting to confront this approach with the use of relative entropy in the context of the holographic correspondence, e.g. \cite{Jafferis15}.

\section*{Aknowledgments}
I would like to thank prof. Nicola Pinamonti for his support and guidance during the writing of this paper, in particular for pointing out a derivation of equation \eqref{flux-AH}.


\begin{thebibliography}{99}
%
\bibitem{HollandsIshibashi19}
S. Hollands and A. Ishibashi. “{News versus information}”. In: \emph{Class. Quant.
Grav.} 36.19 (2019), p. 195001. \textsc{doi}: 10.1088/1361-6382/ab3c1e. arXiv:
1904.00007 \texttt{[gr-qc]}.
%
\bibitem{Longo19}
Roberto Longo. “{Entropy of Coherent Excitations}”. In: \emph{Lett. Math. Phys.} 109.12
(2019), pp. 2587–2600. \textsc{doi}: 10.1007/s11005-019-01196-6. arXiv:
1901.02366 \texttt{[math-ph]}.
%
\bibitem{CasiniGrilloPontiello19}
Horacio Casini, Sergio Grillo, and Diego Pontello. “Relative entropy for coherent states
from Araki formula”. In: \emph{Phys. Rev. D} 99 (12 June 2019), p. 125020. \textsc{doi}:
10.1103/PhysRevD.99.125020. \textsc{url}:
https://link.aps.org/doi/10.1103/PhysRevD.99.125020.
%
\bibitem{Oppenheim15}
Jonathan Oppenheim. “Quantum gravity pioneer”. In: \emph{Nature Physics} 11.10
(Oct. 2015), pp. 805–805. \textsc{issn}: 1745-2481. \textsc{doi}: 10.1038/nphys3499.
\textsc{url}: https://doi.org/10.1038/nphys3499.
%
\bibitem{Misner74}
Charles W. Misner, K.S. Thorne, and J.A. Wheeler. \emph{{Gravitation}}. San Francisco:
W. H. Freeman, 1973. \textsc{isbn}: 978-0-7167-0344-0, 978-0-691-17779-3.
%
\bibitem{Bardeen73}
James M. Bardeen, B. Carter, and S.W. Hawking. “{The Four laws of black hole
mechanics}”. In: \emph{Commun. Math. Phys.} 31 (1973), pp. 161–170. \textsc{doi}:
10.1007/BF01645742.
%
\bibitem{Bekenstein73}
Jacob D. Bekenstein. “Black Holes and Entropy”. In: \emph{Phys. Rev. D} 7 (8 Apr. 1973),
pp. 2333–2346. \textsc{doi}: 10.1103/PhysRevD.7.2333. \textsc{url}:
https://link.aps.org/doi/10.1103/PhysRevD.7.2333.
%
\bibitem{Hawking74BHExplosions}
S.W. Hawking. “{Black hole explosions}”. In: \emph{Nature} 248 (1974), pp. 30–31.
\textsc{doi}: 10.1038/248030a0.
%
\bibitem{Hawking74}
S.W. Hawking. “{Particle Creation by Black Holes}”. In: \emph{Commun. Math. Phys.}
43 (1975). Ed. by G.W. Gibbons and S.W. Hawking. [Erratum: Commun.Math.Phys. 46,
206 (1976)], pp. 199–220. \textsc{doi}: 10.1007/BF02345020.
%
\bibitem{Bombelli86}
Luca Bombelli et al. “{A Quantum Source of Entropy for Black Holes}”. In: \emph{Phys.
Rev. D} 34 (1986), pp. 373–383. \textsc{doi}: 10.1103/PhysRevD.34.373.
%
\bibitem{HollandsSanders17}
Stefan Hollands and Ko Sanders. \emph{{Entanglement measures and their properties
in quantum field theory}}. Feb. 2017. arXiv: 1702.04924 \texttt{[quant-ph]}.
%
\bibitem{Solodukhin11}
Sergey N. Solodukhin. “{Entanglement entropy of black holes}”. In: \emph{Living Rev.
Rel.} 14 (2011), p. 8. \textsc{doi}: 10.12942/lrr-2011-8. arXiv: 1104.3712
\texttt{[hep-th]}.
%
\bibitem{Takesaki70}
M. Takesaki. \emph{{Tomita’s Theory of Modular Hilbert Algebras and its
Applications}}. Vol. 128. Lecture Notes in Mathematics. Springer-Verlag, 1970.
\textsc{doi}: 10.1007/bfb0065832.
%
\bibitem{Haag67}
R. Haag, N.M. Hugenholtz, and M. Winnink. “{On the Equilibrium states in quantum
statistical mechanics}”. In: \emph{Commun. Math. Phys.} 5 (1967), pp. 215–236.
\textsc{doi}: 10.1007/BF01646342.
%
\bibitem{AAQFT15}
Romeo Brunetti et al., eds. \emph{{Advances in algebraic quantum field theory}}.
Mathematical Physics Studies. Springer, 2015. \textsc{isbn}: 978-3-319-21352-1,
978-3-319-21353-8. \textsc{doi}: 10.1007/978-3-319-21353-8.
%
\bibitem{Candelas80}
P. Candelas. “{Vacuum Polarization in Schwarzschild Space-Time}”. In: \emph{Phys.
Rev. D} 21 (1980), pp. 2185–2202. \textsc{doi}: 10.1103/PhysRevD.21.2185.
%
\bibitem{Haag92}
R. Haag. \emph{{Local quantum physics: Fields, particles, algebras}}. Sept. 1992.
\textsc{doi}: 10.1007/978-3-642-61458-3.
%
\bibitem{Kay88}
Bernard S. Kay and Robert M. Wald. “{Theorems on the Uniqueness and Thermal
Properties of Stationary, Nonsingular, Quasifree States on Space-Times with a Bifurcate
Killing Horizon}”. In: \emph{Phys. Rept.} 207 (1991), pp. 49–136. \textsc{doi}:
10.1016/0370-1573(91)90015-E.
%
\bibitem{Kodama79}
Hideo Kodama. “{Conserved Energy Flux for the Spherically Symmetric System and the
Back Reaction Problem in the Black Hole Evaporation}”. In: \emph{Prog. Theor. Phys.}
63 (1980), p. 1217. \textsc{doi}: 10.1143/PTP.63.1217.
%
\bibitem{Hayward93}
S.A. Hayward. “{General laws of black hole dynamics}”. In: \emph{Phys. Rev. D} 49
(1994), pp. 6467–6474. \textsc{doi}: 10.1103/PhysRevD.49.6467.
%
\bibitem{DMP11}
Claudio Dappiaggi, Valter Moretti, and Nicola Pinamonti. “{Rigorous construction and
Hadamard property of the Unruh state in Schwarzschild spacetime}”. In: \emph{Adv.
Theor. Math. Phys.} 15.2 (2011), pp. 355–447. \textsc{doi}:
10.4310/ATMP.2011.v15.n2.a4. arXiv: 0907.1034 \texttt{[gr-qc]}.
%
\bibitem{Hayward97}
Sean A. Hayward. “{Unified first law of black hole dynamics and relativistic
thermodynamics}”. In: \emph{Class. Quant. Grav.} 15 (1998), pp. 3147–3162. \textsc{doi}:
10.1088/0264-9381/15/10/017. arXiv: gr-qc/9710089.
%
\bibitem{Wald84}
Robert M. Wald. \emph{{General Relativity}}. Chicago, USA: Chicago Univ. Pr., 1984.
\textsc{doi}: 10.7208/chicago/9780226870373.001.0001.
%
\bibitem{Abreu10}
Gabriel Abreu and Matt Visser. “{Kodama time: Geometrically preferred foliations of
spherically symmetric spacetimes}”. In: \emph{Phys. Rev. D} 82 (2010), p. 044027.
\textsc{doi}: 10.1103/PhysRevD.82.044027. arXiv: 1004.1456 \texttt{[gr-qc]}.
%
\bibitem{Hayward94}
Sean A. Hayward. “{Gravitational energy in spherical symmetry}”. In: \emph{Phys. Rev.
D} 53 (1996), pp. 1938–1949. \textsc{doi}: 10.1103/PhysRevD.53.1938. arXiv:
gr-qc/9408002.
%
\bibitem{CiolliLongoRuzzi19}
Fabio Ciolli, Roberto Longo, and Giuseppe Ruzzi. “{The information in a wave}”. In:
\emph{Commun. Math. Phys.} (June 2019). \textsc{doi}:
10.1007/s00220-019-03593-3. arXiv: 1906.01707 \texttt{[math-ph]}.
%
\bibitem{HollandsWald14}
Stefan Hollands and Robert M. Wald. “{Quantum fields in curved spacetime}”. In:
\emph{Phys. Rept.} 574 (2015), pp. 1–35. \textsc{doi}:
10.1016/j.physrep.2015.02.001. arXiv: 1401.2026 \texttt{[gr-qc]}.
%
\bibitem{HawkingEllis73}
S.W. Hawking and G.F.R. Ellis. \emph{{The Large Scale Structure of Space-Time}}.
Cambridge Monographs on Mathematical Physics. Cambridge University Press, Feb.
2011. \textsc{isbn}: 978-0-521-20016-5, 978-0-521-09906-6, 978-0-511-82630-6,
978-0-521-09906-6. \textsc{doi}: 10.1017/CBO978051/1524646.
%
\bibitem{Unruh76}
W.G. Unruh. “{Notes on black hole evaporation}”. In: \emph{Phys. Rev. D} 14 (1976),
p. 870. \textsc{doi}: 10.1103/PhysRevD.14.870.
%
\bibitem{Sewell82}
Geoffrey L Sewell. “Quantum fields on manifolds: PCT and gravitationally induced
thermal states”. In: \emph{Annals of Physics} 141.2 (1982), pp. 201–224. \textsc{issn}:
0003-4916. \textsc{doi}: https://doi.org/10.1016/0003-4916(82)90285-8.
\textsc{url}:
https://www.sciencedirect.com/science/article/pii/0003491682902858.
%
\bibitem{Araki76}
H. Araki. “{Relative Entropy of States of Von Neumann Algebras}”. In: \emph{Publ. Res.
Inst. Math. Sci. Kyoto} 1976 (1976), pp. 809–833.
%
\bibitem{Witten18}
Edward Witten. “{APS Medal for Exceptional Achievement in Research: Invited article
on entanglement properties of quantum field theory}”. In: \emph{Rev. Mod. Phys.} 90.4
(2018), p. 045003. \textsc{doi}: 10.1103/RevModPhys.90.045003. arXiv: 1803.04993
\texttt{[hep-th]}.
%
\bibitem{Bisognano75}
J. J Bisognano and E. H. Wichmann. “{On the Duality Condition for a Hermitian Scalar
Field}”. In: \emph{J. Math. Phys.} 16 (1975), pp. 985–1007. \textsc{doi}:
10.1063/1.522605.
%
\bibitem{Brunetti02}
Romeo Brunetti, Daniele Guido, and Roberto Longo. “{Modular localization and Wigner
particles}”. In: \emph{Rev. Math. Phys.} 14 (2002), pp. 759–786. \textsc{doi}:
10.1142/S0129055X02001387. arXiv: math-ph/0203021.
%
\bibitem{KPV21}
Fiona Kurpicz, Nicola Pinamonti, and Rainer Verch. \emph{{Temperature and
entropy-area relation of quantum matter near spherically symmetric outer trapping
horizons}}. Feb. 2021. arXiv: 2102.11547 \texttt{[gr-qc]}.
%
\bibitem{Penrose64}
R. Penrose. “{Conformal treatment of infinity}”. In: (1964). Ed. by C. DeWitt and
B. DeWitt, pp. 565–586. \textsc{doi}: 10.1007/s10714-010-1110-5.
%
\bibitem{Gibbons77}
G. W. Gibbons and S. W. Hawking. “Cosmological event horizons, thermodynamics,
and particle creation”. In: \emph{Phys. Rev. D} 15 (10 May 1977), pp. 2738–2751.
\textsc{doi}: 10.1103/PhysRevD.15.2738. \textsc{url}:
https://link.aps.org/doi/10.1103/PhysRevD.15.2738.
%
\bibitem{Drago17}
Nicolò Drago, Federico Faldino, and Nicola Pinamonti. “{Relative Entropy and Entropy
Production in pAQFT}”. In: \emph{Annales Henri Poincare} 19.11 (2018), pp. 3289–3319.
\textsc{doi}: 10.1007/s00023-018-0730-2. arXiv: 1710.09747 \texttt{[math-ph]}.
%
\bibitem{Jafferis15}
Daniel L. Jafferis et al. “{Relative entropy equals bulk relative entropy}”. In:
\emph{JHEP} 06 (2016), p. 004. \textsc{doi}: 10.1007/JHEP06(2016)004. arXiv:
1512.06431 \texttt{[hep-th]}.
\end{thebibliography}
\end{document}